\documentclass[conference]{IEEEtran}%
\usepackage[linesnumbered,ruled,vlined]{algorithm2e}
\usepackage{graphicx}
\usepackage{bbold}
\usepackage[noend]{algpseudocode}
\usepackage{amsmath,amsbsy,amsthm}
\usepackage{fancyhdr}
\usepackage{enumerate}
\usepackage{array}
\usepackage{verbatim}
\usepackage[table]{xcolor}
\usepackage{subcaption}
\makeatletter
\def\BState{\State\hskip-\ALG@thistlm}
\makeatother

\begin{document}
\title{ LT Codes Combined with Network Coding for Multihop Powerline Smart Grid Networks}
\author{%
{Abraham Kabore{\small }, Vahid Meghdadi{\small }, Jean-Pierre Cances{\small }}%
\vspace{1.6mm}\\
\fontsize{10}{10}\selectfont\itshape
\,Univ. Limoges, CNRS, Xlim UMR 7252, 87000 Limoges, France.\\
\fontsize{9}{9}\selectfont\ttfamily\upshape
\,Email: \{wendyida-abraham.kabore,meghdadi,cances\}@xlim.fr
\vspace{1.2mm}\\
\fontsize{10}{10}\selectfont\rmfamily\itshape
\fontsize{9}{9}\selectfont\ttfamily\upshape}
\maketitle
\begin{abstract}
This paper describes a novel approach for combining Luby Transform (LT) codes and Network Coding (NC) in the context of PowerLine Communications (PLC) smart grid networks. Multihop transmissions of LT-encoded data on PLC networks are considered and algorithms to combine data at relay nodes are proposed. Without the need to decode and then re-encode the total received data stream, the relay nodes can forward the received data stream while adding at the same time their own data. Simulation results are provided confirming the good performance of the proposed algorithms.
\end{abstract}
\section{Introduction}
The goal of the smart grid is to associate to the electricity transmission and distribution networks communication technologies, in order to optimize the production, the transport and the distribution of the electrical energy. 
NarrowBand (NB) PLC, operating below $500$ kHz, is believed to be a natural and cost effective infrastructure choice for smart grid communications \cite{Aalamifar}, as reflected in the adoption of specifications like  G3-PLC, PRIME, and standards like IEEE 1901.2 and ITU-T G.hnem.

The characteristics of the power lines and their mode of operation make the PLC channels non-stationary, highly attenuated and very noisy. To ensure reliable transmissions on PLC channels, the conventional strategies use a fixed rate Forward Error Correction (FEC) code associated with an Automatic Repeat reQuest (ARQ) retransmission mechanism in case the FEC code has failed. 
Since ARQ schemes can lead to a high number of retransmissions and acknowledgments, erasure codes like fountain codes, which exhibit near-optimal overhead, are preferred in replacement of the ARQ mechanism \cite{Byers1,Byers2, Kabore}.
Network coding is attractive for the PLC smart grid networks with the aim of increasing their performance in terms of speed, reliability and efficiency, as demonstrated by the extensive literature on this topic \cite{Phulpin}-\cite{Prior}. For a multipoint-to-point transmission, performing network coding on several LT flows to obtain a larger LT code is not a trivial problem. Indeed, LT codes are extremely sensitive to the statistical degree properties of the encoded packets. Such properties can be altered by a ``naive'' network coding. The problem of distributed LT coding is addressed in this paper for the specific NB PLC smart grid architecture. The association of network coding and LT coding increases the size of the generated LT code and, if optimally combined, the resulting LT code is better in terms of overhead.

\begin{figure}[h]
    \centering
        \includegraphics[width=0.30\textwidth]{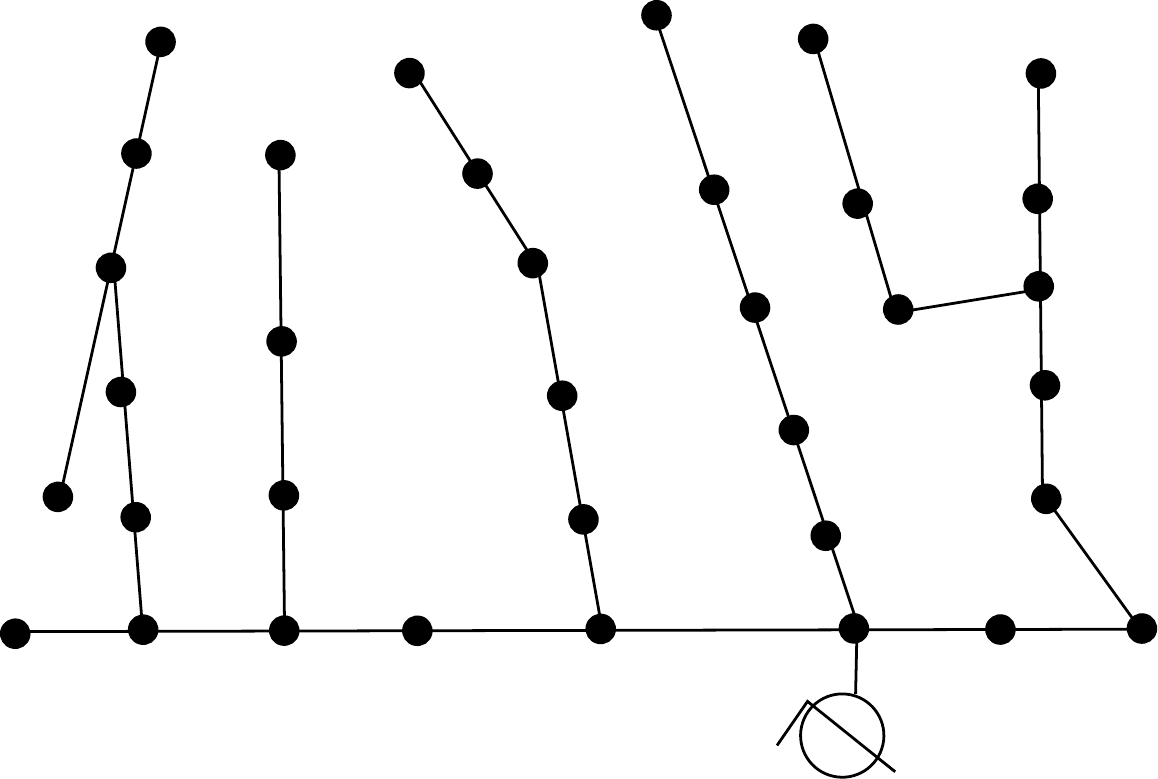}
    \caption{Node test distribution inspired from the IEEE 34 test model.}
    \label{fig:ieee123}
\end{figure}

\begin{figure}[ht]
    \centering
        \includegraphics[width=0.35\textwidth]{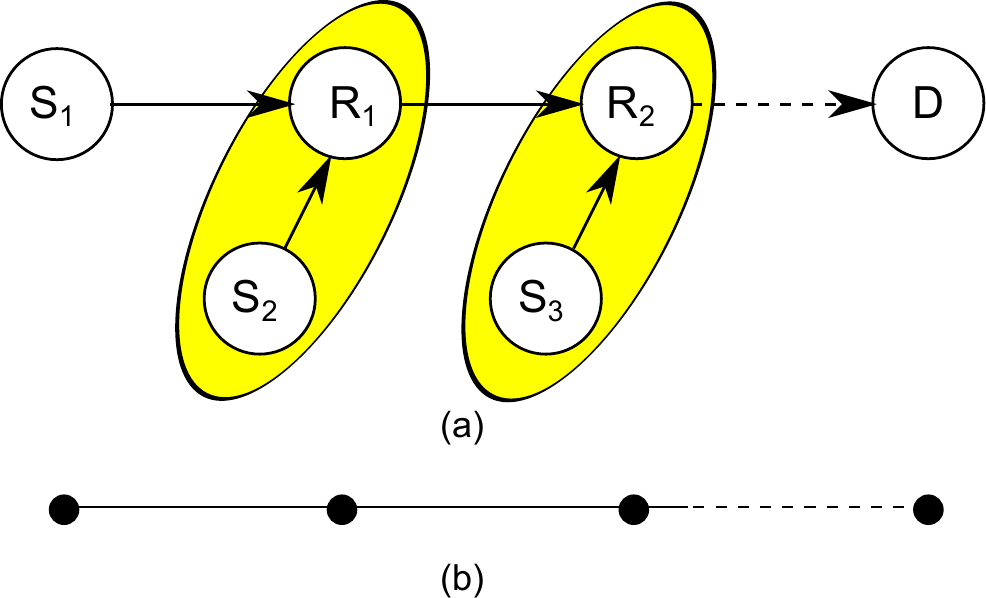}
    \caption{(a) PLC smart grid  data gathering model, (b) equivalent node model presentation as in the IEEE 34 test model}
    \label{fig:schematic}
\end{figure}

The IEEE 34-node power distribution network described in \cite{IEEE123node} and shown in the Figure \ref{fig:ieee123} reveals that the PLC smart grid communication networks presents a tree and radial topology, which is considered in this paper. 
In fact, the network is modeled as a ``line-network'' gathering the information from the leaves (house meters) to the master (concentrator), as presented in the Figure \ref{fig:schematic}-(a). 

When polled, each source node sends its data to the concentrator either directly or through other intermediate source nodes that function as repeaters \cite{Selga}. A repeating function is defined as an integral part of the MAC layer of NB-PLC smart grid networks for the purposes of range extension. 
This type of network is a ``simplified'' version of the well known ``Y networks'' configuration, extensively addressed in the literature \cite{Puducheri}- \cite{Saber}. One major difference from the Y-networks, is that the relay nodes are forwarding the received LT-encoded data stream from a downstream node while adding at the same time their own data i.e, all the packets of the relay nodes are of degree 1.
To the best of our knowledge, this approach has barely been addressed in the literature so far.

The contributions of the paper are the following:
\begin{itemize}
	\item The problem of distributed LT codes is analyzed for the special architecture of NB PLC smart grid networks, where a source node must relay an other source node packets while adding its own packets.
	\item A relaying algorithm based on the conjoint probability on the received degree from S$_1$ and the outgoing degree from the relay is provided. The performance evaluation of the proposed algorithm are presented.
\end{itemize}

The rest of this paper is organized as follows: Section II presents the system and gives a background on LT codes.
Then in Section III, we describe the strategies and algorithms of network coding at the relays that are retained and we expose their design principles.
Section IV presents the simulation results based on the above schemes confirming the system performance. Finally, Section V concludes the paper.
\section{System Model}
Considering the system model of the Figure \ref{fig:ieee123}, one can notice a repeating pattern in the network. 
Most of the configurations encountered in the NB-PLC networks can be summarized by an elementary ``pattern'' consisting of one downstream source node, a bottleneck relay node and the sink node. 
Our set-up is outlined as follows: 
\begin{itemize}
	\item  The downstream node, called S$_1$, transmits an LT encoded message block of length $K_1$. 
	\item  The upstream node, called S$_2$, is collocated with the relay ( $\text{R}_1$).  S$_2$  is also transmitting a message block of length $K_2$ to the sink.
	S$_2$ (S$_1)$ and upstream (downstream) node will be used interchangeably throughout this paper.
	\item  When the sink decodes a particular source, it sends an acknowledgment to that source. The relay stop relaying the discovered packets and the corresponding source stops transmitting.
	\item There is no buffering for the received packets and there is no LT-decoding at the relay nodes. 
\end{itemize}
In this section a short description of LT codes is given \cite{Luby}. 
 Suppose we have to transmit $K$ packets of information. The coded packet $t_n$ is produced from the source block $M$, which comprises $K$ source messages $M = [m_1, m_2, \cdots, m_K]$, by: 
\begin{enumerate}[i)]
	\item Generating a random variable denoted by $d_n$ from the predefined degree distribution.
	\item $t_n$ is obtained by the bitwise sum (mod 2) of $d_n$ packets chosen \emph{uniformly at random} from the $K$ packets.
\end{enumerate}
At the decoder side, upon receiving a coded packet, the belief propagation LT decoder executes the following algorithm:
\begin{enumerate}[i)]
	\item If the packet is of degree 1, the packet is considered discovered. Then all the previously received and all the future packets involving this discovered packet are ``xor-ed'' with the discovered packet. This is to remove its effect and to obtain lower degree packets.
	\item If during the process of step (i) degree 1 packets are generated, repeat the step (i).
	\item repeat step (i) until all the $K$ packets are discovered.
\end{enumerate}

The performance of the LT decoder is very dependent on the degree distribution from which the degrees of the coded packets are chosen. This degree distribution is called output degree distribution. The optimal output degree distribution achieving capacity was found to be the Robust Soliton Distribution (RSD) given by $\mu_K$:
\begin{align}
\mu_K (d)  &=   \frac{\rho(d)+\tau(d)}{(\sum_{i=1}^{K} \rho(i)+\tau(i))},\\
\rho(d) &=    \begin{cases}
    1/K, & \text{if $d=1$}.\\
    1/\big (d \cdot (d-1)\big) & \text{otherwise}.
  \end{cases} and \\
  \tau(d) &=\begin{cases}
    S/K \cdot 1/d, & \text{for $d=1 \cdots \lfloor K/S\rfloor - 1$}.\\
        S \cdot \ln(S/\delta)/K, & \text{for $d= \lfloor K/S \rfloor$}.\\
    0, & \text{otherwise}.
  \end{cases}
\end{align}

Where $K$ is the number of packets to send, $d$ is the degree to be sent, $S = c \cdot \ln (K/\delta)\cdot \sqrt{K}$
, the parameters $c$ and $\delta$ are used to adjust the performance.
In the remainder of this paper, ${\mu}_K$ is a vector of size $K+1$ which denotes an RSD for $K$ source packets. The indices of ${\mu}_K$ start from zero with $ {\mu}_K(0)=0$.

For an optimal belief propagation decoding the packets must be selected at random with uniform distribution during the encoding process. This requirement produces a binomial distribution on the number of edges connected to the check nodes of the bipartite graph representing the LT code, which is called input degree distribution.
 Although it is easy to satisfy this condition for one source, it is quite complicated when the source nodes are distributed and network coding is done in the intermediate nodes. This important issue is considered in this paper.

For a multihop network consisting of two sources, S$_1$ and S$_2$ (S$_2$ being collocated with the relay) with $K_1$, $K_2$ information packets respectively, the objective is to generate an LT code of size $K=K_1+K_2$ preserving the input and output degree distributions.
To respect the input degree distribution at the relay node, each time the relay outputs a coded packet of degree $i$, the probability that this packet comprises $(i-j)$ packets from S$_2$ is calculated as follows:
\[ \frac{{K_1 \choose j} {K_2 \choose i-j}}{{K \choose i}}, \quad 1 \leq i \leq K ,\quad 0\leq j \leq i .
    \label{Eq:binomial}
\]

In order to quantify the impact of a non uniform sampling, some simulations have been carried out. A multipoint-to-point transmission is considered with two source nodes, $K= \{1000, 200\}$, $K_1=K_2=\frac{K}{2}$. The performance in terms of the decoding success rate given a specified redundancy defined as $\epsilon = N/K$ are evaluated, where $N$ is the number of packets sent from the relay before the sink is able to decode. We assume here that all packets of S$_1$ are available at the relay.

\begin{figure*}
\begin{subfigure}{.5\textwidth}
  \centering
  \includegraphics[width=.90\linewidth]{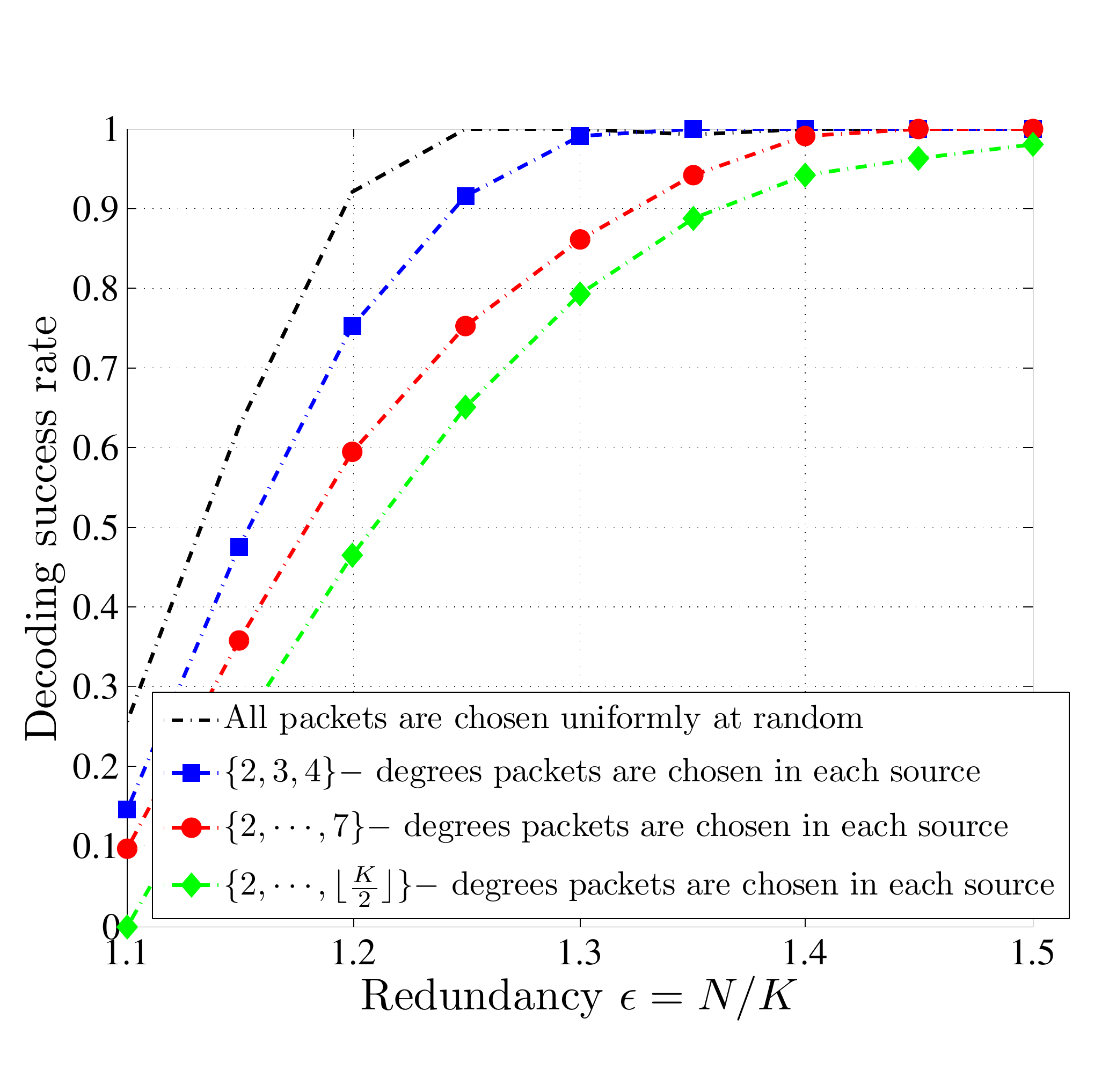}
  \caption{The number of source packets $K = 1000$ and $K_1 = 500$}
  \label{fig:sfig1}
\end{subfigure}%
\begin{subfigure}{.5\textwidth}
  \centering
  \includegraphics[width=.90\linewidth]{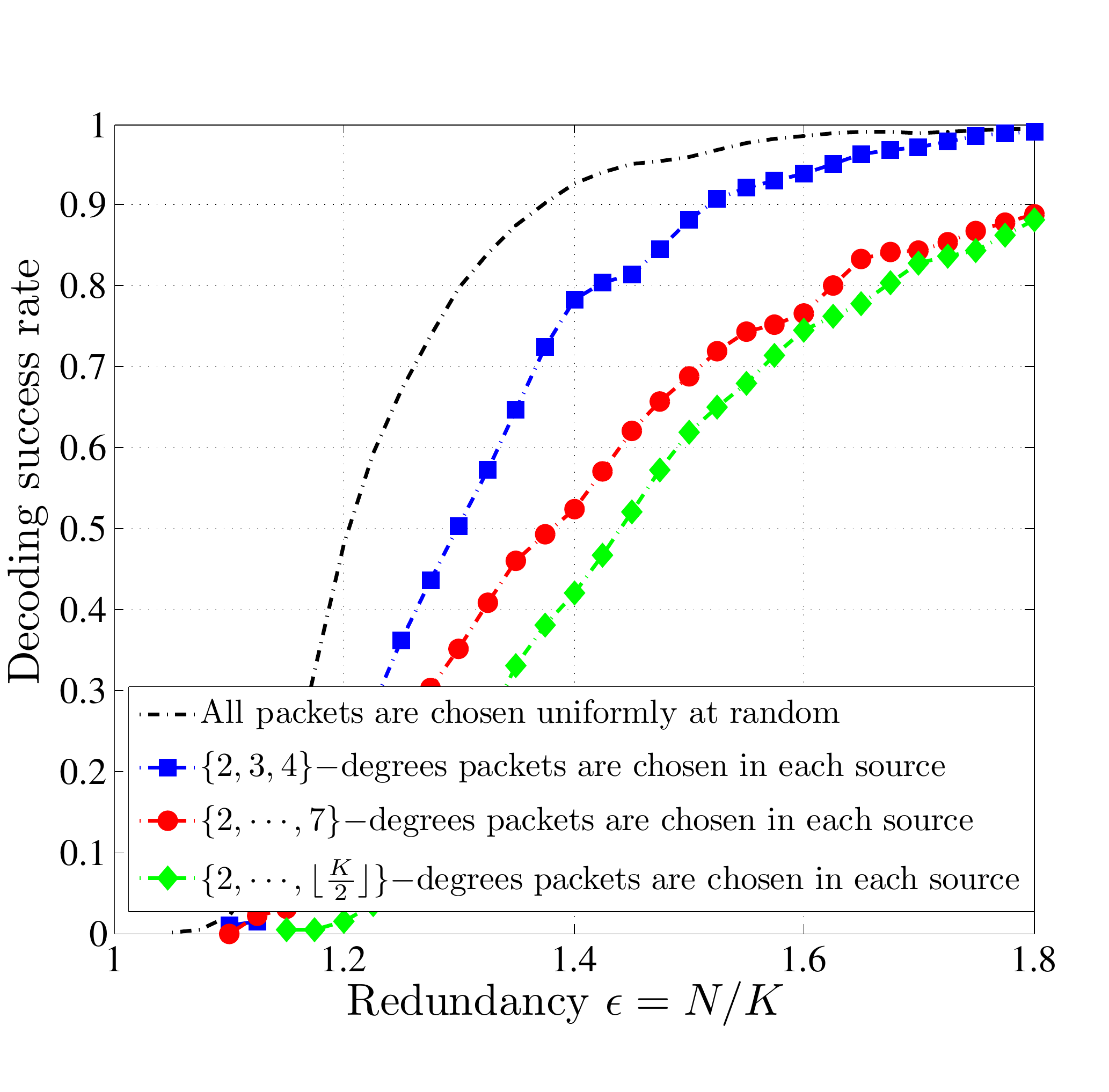}
  \caption{The number of source packets $K = 200$ and $K_1 = 100$}
  \label{fig:sfig2}
\end{subfigure}
\caption{Successful decoding probability of LT codes in terms of overhead when the coded packets are not chosen uniformly at random ($c = 0.05$, $\delta = 0.5$)}
\label{Fig:justification}
\end{figure*}

At first, all the degree 2, 3 and 4 packets are selected exclusively from the source S$_1$ or S$_2$. Therefore a degree 2, 3 or 4 packet is never coded with the packets coming from both sources. The other packets are generated using uniform sampling from the overall packets. Then, the degree 5, 6 and 7 packets are also added to this process and are sampled in the same way as the degrees 2, 3 and 4. And then, all the packets (of degree less than $\lfloor \frac{K}{2} \rfloor$) are selected either in S$_1$ or in S$_2$. Figure \ref{Fig:justification} shows the results. The reference curve in black, corresponds to the ideal case where we have an LT of size $K$.

Using this method to generate the coded packets, redundant coded packets are more likely to appear. This will induce poorer performance of the LT decoding as reflected by the gap between the non uniform sampling curves and the reference curve. This gap is particularly important, when the number of degrees selected in a non uniform manner increases.
From Figure \ref{Fig:justification}, we note that improvements in terms of overhead can be achieved by maintaining a uniform sampling over the set of $K_1+K_2=K$ packets at the relay.

\section{Relaying Strategy}
The objective at the relay is to send packets with an RSD of size $K$ while receiving packets with an RSD of size $K_1$ and having at its disposal $K_2$ packets. Furthermore, the relay should respect the uniform and random selection of the packets resulting in the binomial input degree distribution over all the $K=K_1+K_2$ packets. 
\subsection{Degree distribution management}

Let's define the matrix $\textbf{P}_{(K+1) \cdot (K_1+1)}$, with the entry $p_{i,j}$ being the joint probability that the output packet at the relay is of degree $i$ and comprises $j$ packets from S$_1$. In an ideal case where all the packets are available at the relay the matrix $\textbf{P}$ is computed as:
{\small 
\begin{equation}
\begin{split}
p_{i,j}& = \Pr\{ d=i , d_1=j \}. \\
&= \Pr\{ d=i \} \Pr\{ d_1=j \mid  d=i\}. \\
 &= 
\begin{cases} 
{\mu}_K (i) \frac{{K_1 \choose j}  {K_2 \choose (i-j)}}{{K \choose i}}, & for \quad 0 \leq j \leq i .\\ 
0, &  for \quad i+1 \leq j .
\end{cases}
\end{split}
\label{matrixP}
\end{equation}}

The index begins from zero for notation purposes.
 For example, the 5 first rows and columns of the matrix $\textbf{P}$ for $c=0.05$, $\delta=0.5$ and $K_1 = K_2 = 50$ are computed as follows:
{\small
\begin{equation}
	 \textbf{P}_{101 \times 51} =
\begin{bmatrix}
    0    &  0 & 0 & 0 & 0 & \hdots\\
		0.016 & 0.016 & 0 & 0 & 0 & \hdots\\
		0.11 & \cellcolor{blue!25}0.22 & 0.11 & 0 & 0 & \hdots\\
		0.018 & 0.058 & 0.058 & 0.018 & 0 & \hdots\\
		0.005 & 0.02 & 0.03 & 0.02 & 0.005 & \hdots\\
		\hdots & \hdots & \hdots & \hdots & \hdots & \hdots
\end{bmatrix}
\label{matrixExample}
\end{equation}}
The relay should therefore send $22\%$ packets of degree $2$, formed by ``xor-ing'' one packet from the downstream source and one packet from its own packets.

Since $d$ is the output degree, marginalizing $p(d,d_1)$ with respect to $d_1$ yields
\[P_{\text{out}}(d = i) = \sum_{j = 0}^{K_1} p_{i,j} \quad 0 \leq i \leq K.\] This gives a vector of size $K+1$, obtained as the sum of the columns of the matrix $\textbf{P}$, that is an RSD of size $K$.

The marginal probability $P_\text{\text{S1}}(d_1 = j)$ is a vector of size $ K_1 + 1$ that can be obtained as the sum of the rows of the matrix $\textbf{P}$: 
\[P_{\text{S1}}(j) = \sum_{i = 0}^{K} p_{i,j} \quad 0 \leq j \leq K_1\]

Since the received distribution from S$_1$ is an RSD, it is not always possible to have at the same time a binomial input and an RS output degree distributions at the output of the relay. To better understand this, let us take a look at the matrix given in (\ref{matrixExample}) for example. According to this matrix, the relay needs to output 22\% packets with $d_1=1$ and $d=2$. Since we do not want to decode at the relay, it is simply not possible to have this amount of degree 1 packets from S$_1$ because of the shape of the RSD at S$_1$.

We update $\textbf{P}$ to $\textbf{P}_o$ so that the total need for the degree $j$ packets coming from S$_1$ ($P_\text{S1}(j)$) is less or equal to the available percentage of degree $j$ packets actually coming from S$_1$ (${\mu}_{K_1}(j)$). We therefore need to modify the matrix $\textbf{P}$ to $\textbf{P}_o$, so that the sum of the columns of $\textbf{P}_{o}$ (giving $P_{\text{out}}$) must correspond to an RSD and the sum of the rows of $\textbf{P}_{o}$ (giving $ P_{\text{S1}}$) must always be possible to generate from the received S$_1$ output degree distribution, in other words  $P_{\text{S1}}(j) \leq \mu_{K_1}(j), \, \forall \,j$.

In the following, $\textbf{V}(i,k:n)$ is an index notation referring to the elements $\textbf{V}(i,k), \textbf{V}(i,k+1), \cdots, \textbf{V}(i,n)$ of the matrix $\textbf{V}$. 
\subsection{Matrix $\textbf{P}_o$ computation}
\begin{algorithm}[h]
\DontPrintSemicolon 
\KwIn{ \\$\textbf{P}(i,j)$ \Comment{The ideal joint probability of making a degree $i$ packet at the relay with $j$ packets coming from S$_1$. } \\

${\mu}_{K_1} $ \Comment{ The received output degree distribution from S$_1$}.}
\KwOut{\\
$\textbf{P}_{o}(i,j)$ \Comment{A feasible  joint probability of making a degree $i$ packet at the relay with $j$ packets coming from S$_1$.} \\
}
$\textbf{P}_o = zeros(K+1,K_1 +1)$ \Comment{ Starts $\textbf{P}_o$ to a zero matrix.} \\
$\mu_{K_1} ^{\text{residual}} = {\mu}_{K_1}$ \Comment{Temporary variable  for ${\mu}_{K_1}$.} \\
$\mu_{K_1} ^{\text{residual}} (0) = 1$ \Comment{ Adjust the value of $\mu_{K_1} ^{\text{residual}} (0) $.} \\
\For{$i \gets 1$ \textbf{to} $K$}{

Allocate $\mu_{K_1} ^{\text{residual}} (0 :i)$ to the elements in  $\textbf{P}_o(i,0:i)$ proportionally to the probabilities $\textbf{P}(i,0:i)$ maintaining the condition : $\sum_{j=0}^{j= \min(K_1, i)} \textbf{P}_o(i,j) \leq {\mu}_K (i)$ \\
Update $\mu_{K_1} ^{\text{residual}} (0:i)$  \Comment{Remove the percentage allocated for this $i$th step.} \\
}
\Return {$\textbf{P}_o$}
\caption{Compute $\textbf{P}_o$}
\label{algo:reallocation}
\end{algorithm}
 To calculate $\textbf{P}_o$, the elements of the vector $\mu_{K_1}$ are distributed between the different elements of the matrix $\textbf{P}_{o}$ as follows : 
\begin{itemize}
	 \item As an initialization step, the algorithm starts $\textbf{P}_{o}$ to a zero matrix. $\mu_{K_1}$ is put into a temporary variable ${\mu}_{K_1} ^{\text{residual}}$ and ${\mu}_{K_1} ^{\text{residual}} (0) = 1$.
   \item  For the $i$th row ($i$ starting from 1), the values of $\textbf{P}_o(i, 0:i)$ are increased by distributing ${\mu}_{K_1} ^{\text{residual}}(0:i)$ proportionally to the probabilities $\textbf{P}(i,0:i)$. This allocation is done while respecting the following constraint :  \[\sum_{j=0}^{j= \min(K_1, i)} \textbf{P}_{o}(i,j) \leq {\mu}_K (i).\]
${\mu}_{K_1} ^{\text{residual}}$ is then reduced, on the indexes $0, 1, \cdots, i$, to remove its contribution for this $i$th stage.
\end{itemize}
The main steps of the matrix $\textbf{P}_o$ generation is summarized in algorithm \ref{algo:reallocation}.
We present a trace of the execution of the algorithm \ref{algo:reallocation} with the following parameters $c=0.05$, $\delta=0.5$ and $K_1 = K_2 = 50$. The 4 first rows and columns of the matrix $\textbf{P}_o$ are computed in each iteration. The 4 first rows and columns of the matrix $\textbf{P}$  are computed as follows : 
{\small

 \begin{center}
 \begin{tabular}{ | c || c | c | c| c | c |c | }
		 \hline
		$\textbf{P}(i,j)$ & $j =0$     & $j = 1$    & $j =2$    &   $j =3$       & $\hdots$  \\ \hline
		$i = 0 $              &    0       &      0     &    0      &      0         & $\hdots$  \\ \hline
	  $i = 1 $              &  0.0225    &   0.0225   &    0      &      0         & $\hdots$  \\ \hline
		$i = 2 $              & 0.1099     &    0.2243  &  0.1099   &      0         & $\hdots$  \\ \hline
		$i = 3 $              &   0.0184   &   0.0575   &  0.0575   &    0.0184      & $\hdots$  \\ \hline
		$\vdots$              &   $\vdots$ &   $\vdots$ &  $\vdots$ &   $\vdots$     & $\vdots$  \\ \hline
    $\hdots$              &   $\hdots$ &   $\hdots$ &  $\hdots$ &   $\hdots$     & $\hdots$  \\ \hline
  \end{tabular} 
\end{center} }

$\textbf{P}_o(i,j)$ is initialized to a zero matrix  and $\mu_{K_1} ^{\text{residual}} = \mu_{K_1}$.

{\small
\begin{center}
 \begin{tabular}{ | c || c | c | c| c | c |c | }
    \hline
		$\textbf{P}_{o}(i,j)$ & $j =0$     & $j = 1$    & $j =2$    &   $j =3$       & $\hdots$  \\ \hline
		$i = 0 $              &    0       &      0     &    0      &      0         & $\hdots$  \\ \hline
		$i = 1 $              &    0       &      0     &    0      &      0         & $\hdots$  \\ \hline
		$i = 2 $              &    0       &      0     &    0      &      0         & $\hdots$  \\ \hline
		$i = 3 $              &    0       &      0     &    0      &      0         & $\hdots$  \\ \hline
		$\vdots$              &   $\vdots$ &   $\vdots$ &  $\vdots$ &   $\vdots$     & $\vdots$  \\ \hline
    $\hdots$              &   $\hdots$ &   $\hdots$ &  $\hdots$ &   $\hdots$     & $\hdots$  \\ \hline
		\hline 
		\hline 
		$\mu_{K_1} ^{\text{residual}}$ & 1 &  0.0316    &   0.4442   &     0.1519    &  $\hdots$  \\ \hline
  \end{tabular}
\end{center} }

Given that it is always possible to generate packets of degree $d_1 =0$, we arbitrarily set \[ \mu_{K_1} ^{\text{residual}} (0) = 1 ;\] 
$\mu_{K_1} ^{\text{residual}} (0) $ being the percentage of packets of degree $ d_1 = 0 $, coming from S$_1$, employed in the merging process.

In the first iteration, we allocate $\mu_{K_1}^{\text{residual}}(0: 1)$ to the elements in $\textbf{P}_o(1,0 : 1)$. 
The allocation is done in such a way as to best maintain the percentages found in $\textbf{P} (1, 0 : 1)$. $\mu_{K_1} ^{\text{residual}}$ is updated on the indexes $(0:1)$.

 {\small
 \begin{center}
 \begin{tabular}{ | c || c | c | c| c | c |c | }
		 \hline
		$\textbf{P}_{o}(i,j)$ & $j =0$     & $j = 1$    & $j =2$    &   $j =3$       & $\hdots$  \\ \hline
		$i = 0 $              &    0       &      0     &    0      &      0         & $\hdots$  \\ \hline
		$i = 1 $              &\cellcolor{blue!25} 0.0225  &  \cellcolor{blue!25}0.0225    &    0      &      0         & $\hdots$  \\ \hline
		$i = 2 $              &    0       &      0     &    0      &      0         & $\hdots$  \\ \hline
		$i = 3 $              &    0       &      0     &    0      &      0         & $\hdots$  \\ \hline
		$\vdots$              &   $\vdots$ &   $\vdots$ &  $\vdots$ &   $\vdots$     & $\vdots$  \\ \hline
    $\hdots$              &   $\hdots$ &   $\hdots$ &  $\hdots$ &    $\hdots$    & $\hdots$  \\ \hline
		\hline 
		\hline 
		$\mu_{K_1} ^{\text{residual}}$ & \cellcolor{blue!25} 0.9775 & \cellcolor{blue!25} 0.091    &   0.4442   &     0.1519    &  $\hdots$  \\ \hline
		
  \end{tabular}

\end{center} 
}

In the second iteration, we allocate $\mu_{K_1}^{\text{residual}}(0: 2)$ to the elements in $\textbf{P}_o(2,0 : 2)$. 
As in the previous iteration, the allocation is done in such a way as to best maintain the percentages found in $\textbf{P} (2, 0 : 2)$.
 $\mu_{K_1} ^{\text{residual}}$ is updated on the indexes $(0: 2)$ afterwards.

 {\small
 \begin{center}
 \begin{tabular}{ | c || c | c | c| c | c |c | }
		 \hline
		$\textbf{P}_{o}(i,j)$ & $j =0$     & $j = 1$    & $j =2$    &   $j =3$       & $\hdots$  \\ \hline
		$i = 0 $              &    0       &      0     &    0      &      0         & $\hdots$  \\ \hline
		$i = 1 $              &    0.016   &     0.016  &    0      &      0         & $\hdots$  \\ \hline
		$i = 2 $              & \cellcolor{blue!25}   0.1766  &    \cellcolor{blue!25} 0.091  &   \cellcolor{blue!25} 0.1766  &      0         & $\hdots$  \\ \hline
		$i = 3 $              &    0       &     0       &    0       &    0         & $\hdots$  \\ \hline
		$\vdots$              &   $\vdots$ &   $\vdots$ &  $\vdots$ &   $\vdots$     & $\vdots$  \\ \hline
    $\hdots$              &   $\hdots$ &   $\hdots$ &  $\hdots$ &    $\hdots$    & $\hdots$  \\ \hline
		\hline 
		\hline 
		$\mu_{K_1} ^{\text{residual}}$& \cellcolor{blue!25}  0.8009 &  \cellcolor{blue!25} 0    &   \cellcolor{blue!25} 0.2654   &     0.1519    &  $\hdots$  \\ \hline
		
  \end{tabular}
\end{center}
}
In the third iteration, we allocate $\mu_{K_1}^{\text{residual}}(0: 3)$ to the elements in $\textbf{P}_o(3,0 : 3)$ according to the same process as above.
 $\mu_{K_1} ^{\text{residual}}$ is updated accordingly. 
{\small
 \begin{center}
 \begin{tabular}{ | c || c | c | c| c | c |c | }
		 \hline
		$\textbf{P}_{o}(i,j)$ & $j =0$     & $j = 1$    & $j =2$    &   $j =3$       & $\hdots$  \\ \hline
		$i = 0 $              &    0       &      0     &    0      &      0         & $\hdots$  \\ \hline
		$i = 1 $              &    0.016   &     0.016  &    0      &      0         & $\hdots$  \\ \hline
		$i = 2 $              &   0.1766   &   0.091   &    0.1766  &      0         & $\hdots$  \\ \hline
		$i = 3 $              &   \cellcolor{blue!25}  0.0195      &  \cellcolor{blue!25}    0      &    \cellcolor{blue!25}  0.0610    &  \cellcolor{blue!25}   0.0195         & $\hdots$  \\ \hline
		$\vdots$              &   $\vdots$ &   $\vdots$ &  $\vdots$ &   $\vdots$     & $\vdots$  \\ \hline
    $\hdots$              &   $\hdots$ &   $\hdots$ &  $\hdots$ &    $\hdots$    & $\hdots$  \\ \hline
		\hline 
		\hline 
		$\mu_{K_1} ^{\text{residual}}$& \cellcolor{blue!25} 0.7814&   \cellcolor{blue!25} 0    &  \cellcolor{blue!25}  0.2044   &  \cellcolor{blue!25}    0.1324    &  $\hdots$  \\ \hline
		
  \end{tabular}
\end{center}}

The update is done the same way for the $i$th iteration, up to $i =K$.
\subsection{Output degree generation}
Whenever a degree $d_1 = j$ is received from S$_1$, the relay has two options in order to generate the output degree $d = i$.
\begin{itemize}
	\item  Either it consults the column with the index $j$ of the matrix $\textbf{P}_{o}$ and thus forms a \emph{pmf} (probability  mass function) by normalizing this column to make it a valid \emph{pmf}. $d$ is drawn from this \emph{pmf}. It then outputs a degree $d$ packet which comprises the received packet and $d- j$ of its own source packets sampled at random.
  \item Or, it selects to output a coded packet including exclusively its own packets.
	\end{itemize}
 
One notes that the relay never receives $d_1=0$ packets. Therefore, in order to generate the packets coming exclusively from its own set of packets, which happens with probability $P_{\text{exc-S2}}=P_{\text{S1}}(0)$, the relay operates as follows. Upon receiving a packet from S$_1$ of degree $d_1=j$ that occurs with the probability ${\mu}_{K_1}(j)$, at $P_{\text{S1}}(j) / {\mu}_{K_1}(j)$ of the time the received packet is mixed with packets of S$_2$ to form the output packet as described before. And for the rest of the time, the column of indice zero of the matrix 
$\textbf{P}_{o}$ is used to produce packets coming exclusively from S$_2$. The percentage of the packets coming exclusively from S$_2$ will be the same as $P_{\text{S1}}(0)$ while the output degree distribution stays RS.

 The detail of the procedure is given in the pseudo code of algorithm \ref{algo:merge}.

\begin{algorithm}[ht]
\DontPrintSemicolon 
\KwIn{ \\
 $\textbf{P}_o(d,d_1)$ \Comment{The joint probability of making a degree $d$ coded packet at the relay with $d_1$ packets coming from S$_1$.}\\
 Pck$_{in}$ \Comment{The received packet from S$_1$.}\\
${\mu}_{K_1} $ \Comment{The received output degree distribution from S$_1$.} }
\KwOut{\\ 
 Pck$_{out}$ \Comment{The generated packet to be sent to the sink.}} 
    $j = \text{degree} (\text{Pck}_{in})$ \Comment{ The received degree.} \\
		$P_{\text{S1}}(j) = \sum_{i = 0}^{K}\textbf{P}_o(i, j)$.\\
    $P_{\text{using}}(j) = P_{\text{S1}}(j) / {\mu}_{K_1}  (j)$ \Comment{ The probability of using $\text{Pck}_{in}$ in the merging process.}\\
        \If{$ rand \leq P_{\text{using}}(j) $ }{ 
				Form a \emph{pmf}: $\Pr(D = d)= \textbf{P}_o(d,j)/ \big (\sum_{d=0}^{K}\textbf{P}_o(d,j)\Big )$ for the choice of the degree $d$.\\
				Choose $d$ by sampling the $\emph{pmf}$.\\
				Sample at random $d - j$  packets from the relay to form Pck$_r$.  \\
				Send $\text{Pck}_{out}= \text{Pck}_{in} \oplus \text{Pck}_r $.\\}
       \Else{Choose a degree $d$ according to the \emph{pmf}: $ \Pr(D = d) = \textbf{P}_o(d,0)/ \big (\sum_{d=0}^{K}\textbf{P}_o(d,0)\Big )$.\\
             Sample at random $d$ packets from the relay to form $\text{Pck}_{out}$. }
\caption{ Packets merging algorithm}
\label{algo:merge}
\end{algorithm}

\section{Simulation results}  
In this section, simulation results are presented for the proposed algorithms and are essentially compared with two simulation scenarios: 
\begin{itemize}
	\item a \textit{standard LT} i.e. we consider a point-to-point transmission with only one source that transmits $(K_1 + K_2)$ LT-encoded packets to the sink.
   \item a time-multiplexing of two LT codes i.e. the relay alternately sends the LT-encoded packets coming from S$_1$ and its own LT-encoded packets.
\end{itemize}
We evaluate the performance in terms of the decoding success rate given a specified redundancy defined as $\epsilon = N/K$, where $N$ is the number of packets sent from the relay before the sink is able to decode.

\begin{figure}[ht]
\begin{flushleft}
        \includegraphics[width=0.5\textwidth]{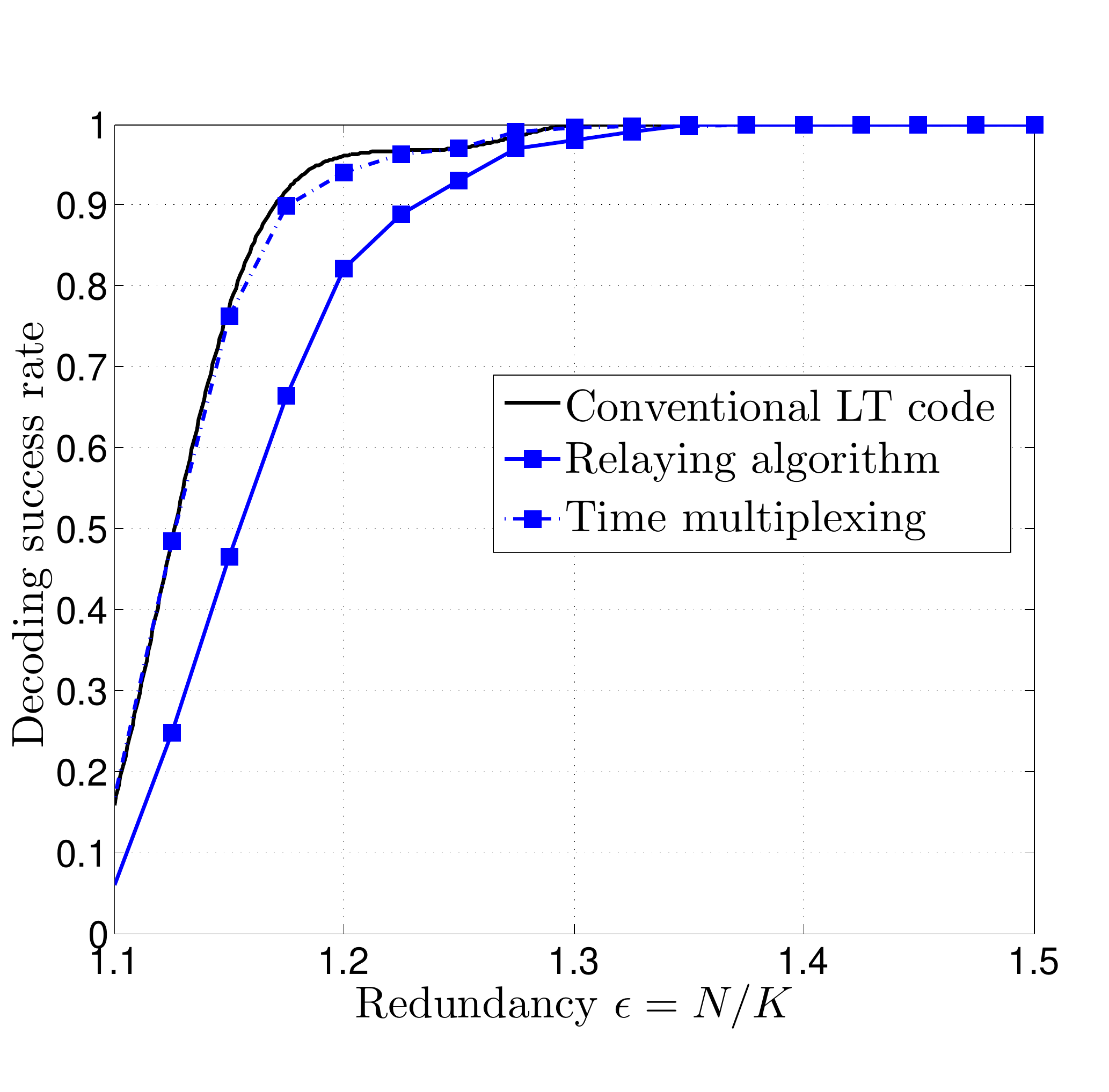}
\end{flushleft}
    \caption{Successful decoding probability of the merging process in terms of overhead (with $K1 =K2 =150$, $c = 0.05$,$\delta = 0.5$)}
    \label{Fig:MergingProcess1}
\end{figure}
Figure \ref{Fig:MergingProcess1} shows the performance of the proposed method compared with the standard LT and the time-multiplexing forwarding scheme for $K = 1000$ and $K_1 = 500$. 
It is observed that, the proposed merging process outperforms the time-multiplexing forwarding scheme, especially when the overhead increases. For example, for a $90\%$ successful decoding rate, the proposed algorithm requires about $5\%$ less overhead than the alternative time-multiplexing LT coding.
\begin{figure}[ht]
\begin{flushleft}
        \includegraphics[width=0.5\textwidth]{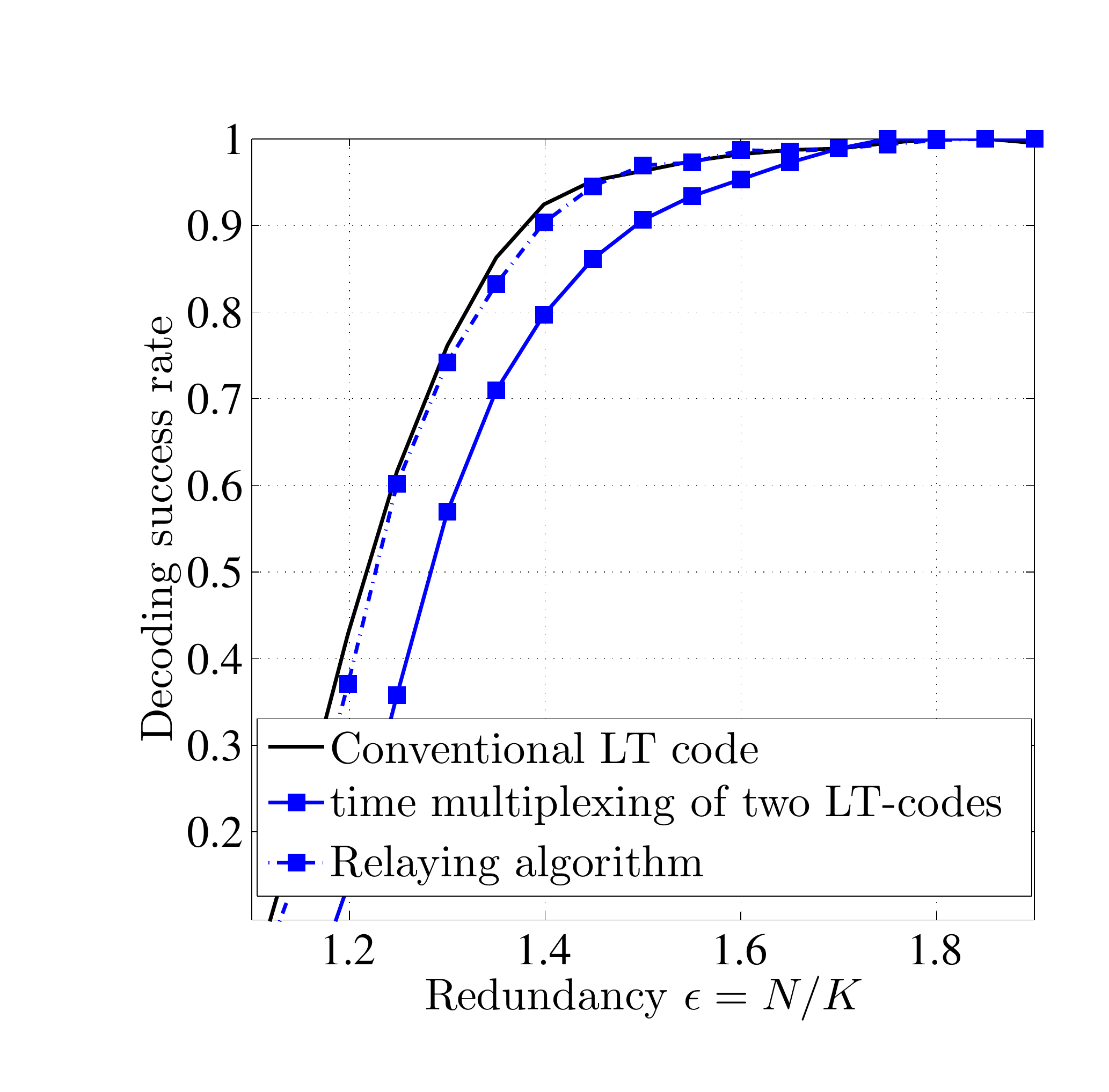}
\end{flushleft}
    \caption{Successful decoding probability of the merging process in terms of overhead (with $K1 =K2 =150$, $c = 0.05$, $\delta = 0.5$)}
    \label{Fig:MergingProcess2}
\end{figure}
For a smaller size of $K$ ($K = 200$, $K_1 = 100 $), the gap between the time-multiplexing approach and the proposed algorithm is more important, as described in Figure \ref{Fig:MergingProcess2}. In this Figure, one can see that the curve corresponding to the standard LT code is extremely closed to the curve corresponding to the proposed relaying algorithm. In addition, the gap between the performance of the time multiplexing and the performance of the proposed algorithm is wider; e.g. for $90\%$successful decoding percentage, the proposed algorithm requires about $10\%$ less overhead than the alternative time-multiplexing LT coding.
%

\section{Conclusion}
This paper presents a strategy of relaying fountain code while using inter-session network coding for the special topology of NB-PLC networks for smart grid applications. We consider the relaying of fountain coded packets on a multihop transmission link and we propose algorithms to combine the packets at the relay, in a way to preserve the important properties that allow optimal decoding at the sink. Simulation results confirm the good performance of the proposed algorithm for a realistic network. 

\end{document}